\documentclass{article}

\usepackage{arxiv}

\usepackage[utf8]{inputenc} 
\usepackage[T1]{fontenc}    
\usepackage{hyperref}       
\usepackage{url}            
\usepackage{booktabs}       
\usepackage{amsfonts}       
\usepackage{nicefrac}       
\usepackage{microtype}      
\usepackage{lipsum}		
\usepackage{graphicx}
\usepackage{doi}
\usepackage{url} 
\usepackage{multirow}
\usepackage{float}
\usepackage[utf8]{inputenc}
\usepackage{amsmath}
\usepackage{amssymb}
\usepackage{array}
\usepackage{tabularx}
\usepackage{verbatim}
\usepackage{xcolor}
\usepackage[linesnumbered,ruled]{algorithm2e}
\usepackage{subcaption}
\usepackage[noend]{algpseudocode}
\usepackage{blindtext}
\usepackage[normalem]{ulem}
\usepackage{ragged2e}
\usepackage{paralist}
\usepackage{subfiles} 
\SetKwInput{KwInput}{Input} 
\SetKwInput{KwOutput}{Output}
\usepackage{svg}
\usepackage{enumitem}

\title{Single Event Transient Fault Analysis of ELEPHANT cipher}

\author{ \href{https://orcid.org/0000-0002-1708-8462}{\includegraphics[scale=0.06]{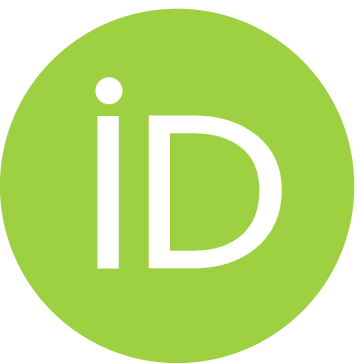}\hspace{1mm}Priyanka Joshi}\\
	Department of Computer Science\\
	Indian Institute of Technology Indore\\
	India, 453552 \\
	\texttt{phd1801201001@iiti.ac.in} \\
	\and
	\href{https://orcid.org/0000-0003-1883-4639}{\includegraphics[scale=0.06]{orcid.eps}\hspace{1mm}Bodhistwa Mazumdar}\\
	Department of Computer Science\\
	Indian Institute of Technology Indore\\
	India, 453552 \\
	\texttt{bodhistwa@iiti.ac.in} \\
}

\renewcommand{\shorttitle}{\textit{arXiv} Template}

\hypersetup{
pdftitle={Fault analysis of ELEPHANT},
pdfsubject={},
pdfauthor={Priyanka Joshi, Bodhisatwa Mazumdar},
pdfkeywords={Fault Analysis, Elephant, authenticated encryption},
}

\begin{document}
\maketitle

\begin{abstract}
In this paper, we propose a novel fault attack termed as {\em Single Event Transient Fault Analysis } (SETFA) attack, which is well suited for hardware implementations. The proposed approach pinpoints hotspots in the cipher's Sbox combinational logic circuit that significantly reduce the key entropy when subjected to faults. ELEPHANT is a parallel authenticated encryption and associated data (AEAD) scheme targeted to hardware implementations, a finalist in the Lightweight cryptography (LWC) competition launched by NIST. In this work, we investigate vulnerabilities of ELEPHANT against fault analysis. We observe that the use of 128-bit random nonce makes it resistant against many cryptanalysis techniques like differential, linear, etc., and their variants. However, the relaxed nature of Statistical Fault Analysis (SFA) methods makes them widely applicable in restrictive environments. We propose an SETFA-based key recovery attack on {\em Elephant}. We performed Single experiments with random plaintexts and keys, on {\em Dumbo}, a Spongent-based instance of {\em Elephant-AEAD} scheme. Our proposed approach could recover the secret key in $85-250$ ciphertexts. In essence, this work investigates new vulnerabilities towards fault analysis that may require to be addressed to ensure secure computations and communications in IoT scenarios.
\end{abstract}
\section{Objective}
Sbox is the most decisive component of any cryptographic algorithm. We aim at pinpointing the most vulnerable hotspots in a Sbox circuit against a class of fault attacks in this work. ELEPHANT cipher is designed to operate efficiently and securely in highly constrained environments like the Internet of Things (IoT), where fault attacks make a potent threat. This work intends to evaluate the security of ELEPHANT against a class of fault analysis attacks.
\section{Elephant cipher}
Elephant parallel authenticated encryption scheme, designed by Beyne et al. \cite{DBLP:journals/tosc/BeyneCDM20}, is a round-2 candidate in the NIST lightweight cryptography (NIST-LWC) standardization process. The designers proposed three variants under the Elephant AEAD scheme, {\em Dumbo}, {\em Jumbo}, and {\em Derilium}. Dumbo and Jumbo are targeted for hardware implementations, and are based on {\em Spongent} \cite{DBLP:conf/ches/BogdanovKLTVV11} permutation with $160$-bit and $176$-bit block sizes, respectively. {\em Derilium} is based on {\em Keccak} \cite{DBLP:conf/dagstuhl/BertoniDPA09a} with $200$-bit block size, and it is targeted for software implementations. In this work, we consider the {\em Dumbo} variant of Elephant-AEAD scheme. However, the analysis provided for {\em Dumbo} is also applicable on {\em Jumbo} with a minor change in the size of the permutation. The description of {\em Dumbo} is provided below.
\vskip 0.1in
\paragraph{\large Dumbo:}
\begin{figure}[!ht]
	\centering
	\includegraphics[width = 9cm]{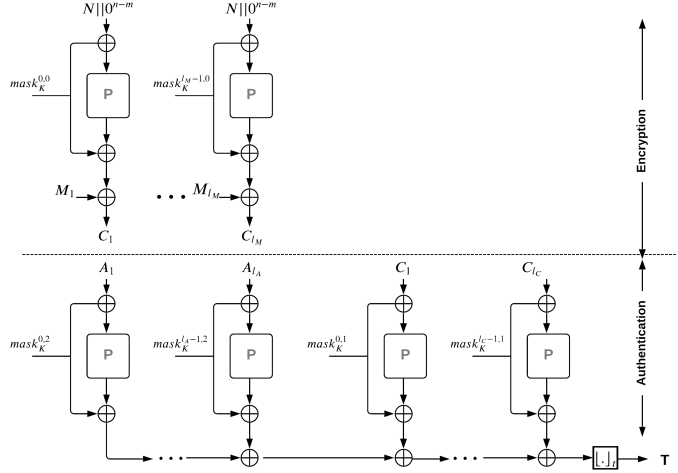}
	\caption{Illustration of Elephant-AEAD scheme \cite{DBLP:journals/tosc/BeyneCDM20}}
	\label{fig:elephant}
\end{figure}

The Elephant-AEAD scheme is illustrated in the Figure \ref{fig:elephant}. $M = M_1,M_2,\dots, M_{l_M}$ denotes message, where $M_i$ is $160$-bit block,  $A = A_1,A_2,\ldots, A_{l_A}$ denotes associated data, and $C = C_1,C_2,\ldots, C_{l_c}$ is corresponding ciphertext. $N$ is a $96$-bit input, called {\em nonce}, $T$ is $64$-bit output, called {\em Tag} used for authentication, and $K$ is a $128$-bit secret key. {\em Dumbo} is primary member of NIST-submission with state size of $160$ bits.

In the figure, $P$ is a $160$-bit {\em Spongent} permutation, and {\em mask} is a word-LFSR based masking function \cite{DBLP:conf/eurocrypt/GrangerJMN16}. The function {\em mask} is implemented using two functions $\varphi_1$ and $\varphi_2$. $\varphi_1$ : $\{0,1\}^n \rightarrow \{0,1\}^n$ is an $n$-bit LFSR (value of $n$ is $160$ for {\em Dumbo}). $\varphi_2 = \varphi_1 \oplus id$, where $id$ is the identity function. The function {\bf {\em mask}} : $\{0,1\}^k \times  \mathbb{N}^2 \rightarrow \{0,1\}^n$ is defined in Equation~\ref{eq:mask}. The LFSR $\varphi_1$ is implemented as shown in Equations below.
\begin{eqnarray}
	mask^{0,0}_K = mask(K,a,b) = {\varphi_2}^a \circ {\varphi_1}^b \circ  P(K||0^{n-k})\label{eq:mask}\\
	x_0 \leftarrow x_0\lll3\oplus x_3\ll7 \oplus x_{13}\gg7 \nonumber\\
	(x_0,\cdots,x_{19}) \mapsto (x_1,\cdots,x_{19},x_0)
	\label{eq:lfsr1}
\end{eqnarray}
\subsection{Encryption}
As depicted in Figure \ref{fig:elephant}, there are four inputs to the encryption operation: a key $K \in \{0,1\}^k$, a nonce $N \in \{ 0,1\}^m$, a message $M \in \{0,1\}^*$, and associated data $A \in \{0,1\}^*$. The output of the encryption is a ciphertext $C \in \{0,1\}^{|M|}$, and a tag $T \in \{0,1\}^t$. For {\em Dumbo}, the values of $k$, $m$, $n$, and $t$ are $128$, $96$, $160$, and $64$, respectively. As shown in Figure \ref{fig:elephant}, the encryption phase of {\em Elephant} cipher scheme works in parallel mode to process arbitrary length input message $M$. Each parallel branch processes $n$-bit message block. In each parallel branch, the $k$-bit key $K$ is processed using {\em mask} function with branch specific parameters $a$, and $b$. The $n$-bit output of {\em mask} function, called expanded key, $K'$, is then used as a key for encryption in that branch. The $m$-bit nonce $N$ is padded with $zeros$ and passed to permutation $P$ after being $XOR$ed with $K'$. The output of $P$ is again $XOR$ed with $K'$, which is then $XOR$ed with $i^{th}$ message block $M_i$. Similarly, in authentication phase associated data is processed along with ciphertexts generated in encryption phase to produce $t$-bit authentication tag $T$.
\subsection{ Decryption}
The decryption algorithm takes $five$ inputs: a key $K \in \{0,1\}^k$, a nonce $N \in \{ 0,1\}^m$, a ciphertext $C \in \{0,1\}^{|M|}$, a tag $T \in \{0,1\}^t$, and associated data $A \in \{0,1\}^*$. The output of the decryption is a message $M \in \{0,1\}^*$ if the tag $T$ is verified, otherwise, the algorithm returns a special symbol $\perp$.
\subsection{Spongent Permutation}
Spongent-$160$ is used as permutation in Dumbo authenticated encryption. The permutation applies round operations on the $160$-bit input $X$ iteratively for $80$ rounds. The round operations are as follows.
\begin{enumerate}
	\item {\em Constant addition:}
	A round constant is generated using a $7$-bit LFSR called {\bf{\em iCounter}} defined by the primitive polynomial $p(x) = x^7 + x^6 + 1$, with initial value $1000101$, then it is added to the round input $X$ as follows:
	\begin{eqnarray}
		Rc \leftarrow 0^{153}||iCounter_{160}(i)\nonumber \\
		Rev_{Rc} \leftarrow rev(Rc) \nonumber \\
		X \leftarrow X \oplus Rc \oplus Rev_{Rc} \nonumber
	\end{eqnarray}
	The $rev$ function reverses the order of the bits of its input.
	\item {\em sBoxLayer:} A $4$-bit Sbox is applied $40$ times in parallel. The Sbox mapping is shown is following table.
	\begin{table}[H]
		\centering
		\begin{tabular}{|p{0.4cm}|p{0.06cm}|p{0.06cm}|p{0.06cm}|p{0.06cm}|p{0.06cm}|p{0.06cm}|p{0.06cm}|p{0.06cm}|p{0.06cm}|p{0.06cm}|p{0.06cm}|p{0.06cm}|p{0.06cm}|p{0.06cm}|p{0.06cm}|p{0.06cm}|}
			\hline
			$x$&0&1&2&3&4&5&6&7&8&9&A&B&C&D&E&F\\
			\hline
			$S(x)$&E&D&B&0&2&1&4&F&7&A&8&5&9&C&3&6\\
			\hline
		\end{tabular}
		\caption{Spongent Sbox}
		\label{tab:spongentSbox}
	\end{table}
	\item {\em pLayer:} This layer shuffles the input bits using the bit-permutation, $P_{160}(j)$.
	\begin{equation}
		P_{160}(j) =  \left\{\begin{matrix}
			40\hspace{0.05cm}\cdot  \hspace{0.05cm} j \hspace{0.1cm}mod \hspace{0.1cm}159, & if \hspace{0.05cm} j\in\{0,\dots,158\},\\ 
			\hspace{-1.8cm}159 & \hspace{-1.1cm}if \hspace{0.05cm}j\hspace{0.1cm}=\hspace{0.1cm}159.
		\end{matrix}\right. \nonumber
	\end{equation}
\end{enumerate}

\section{Threat Model}
For the proposed {\em Single Event Transient Fault Analysis} (SETFA) method, we assume that the targeted S-box in the block cipher is implemented as a combinational logic circuit. The adversary is capable to inject Single Event Transient (SET) a special case of transient stuck-at faults\cite{DBLP:journals/tc/Leveugle07} at specific intermediate wires in the Sbox circuit. Once an SET fault is injected, the attacker can collect sufficient number of faulty ciphertexts. We assume that separate circuits of Spongent permutation are used for computing mask and ciphertext. The adversary can utilize either {\em Chosen Plaintext Attack} (CPA) and/or {\em Known Plaintext Attack} (KPA) attack models.

\section{Proposed Attack}

\subsection{SETFA}
The proposed {\em Single Event Transient Fault Analysis} (SETFA) is performed in five phases as follows: 
\begin{enumerate}
    \item[(i)] {\bf Indentify hot-spots:} Hot-spots are the fault points which, when subjected to a particular fault combination, alter the probability distribution of Sbox outputs. 
    \item[(ii)] {\bf Choose a fault combination:} Attacker chooses a suitable fault combination for injecting faults.
    \item [(iii)]{\bf Fault injection:} The targeted cryptographic implementation is subjected to faults specific to the fault combination chosen in previous step.
    \item[(iv)] {\bf Encrypt with faults:} Execute the encryption function and collect faulty ciphertexts for analysis.
    \item [(v)]{\bf Fault Analysis:} A statistical analysis technique is used to recover the key. 
\end{enumerate}

\subsection{SETFA on Dumbo- Elephant cipher}
An attacker can use the proposed {\em SETFA} method to recover the secret key of Dumbo cipher using the below-mentioned steps.
\begin{itemize}
    \item Identify hot-spots in Spongent Sbox.
    \item Choose optimal fault combination for fault injection. For experiment purposes, we chose fault combinations that lead to minimum residual key-space. In other words, fault combinations that result in single missing value in faulty Sbox output. 
    \item Inject the required {\em SET} faults for the chosen fault combination.
    \item Collect the faulty ciphertext $C'$.
    \item Compute faulty intermediate text using ${I'}_1 = {\left \lceil{C'}\right \rceil}_{160} \oplus M_1$, where ${\left \lceil{C'}\right \rceil}_{160}$ . As per the fault model, $M_1$ being the first plaintext message which is known to the attacker. 
    \item Apply probability distribution based statistical analysis on ${I'}_1$ to recover the expanded key $K'$.
    \item Recover the key $K$ from $K'$ by inverting Spongent permutation.
\end{itemize}
\section{Experiments and results}
We used simulation in C language for the validation of the proposed key recovery attack. The equations corresponding to the Spongent Sbox are as follows. The $4$-bit input and output of the Sbox are denoted as $X = X_0X_1X_2X_3$, and $S(X) = Y_0Y_1Y_2Y_3$, where $X_0$ and $Y_0$ are considered as MSBs.

\begin{footnotesize}
\begin{eqnarray}
  \left.\begin{aligned}
    Y_0 &= \overline{( X_0\oplus  X_1) +X_2)} + \overline{(X_1 + (X_2  \odot X_3))}+\overline{\overline{(\overline{X_0}\cdot X_1)} + \overline{(X_2 \cdot X_3)}}\nonumber\\
    Y_1 &= \overline{X_0 + (X_1 \oplus X_2)} + \overline{X_1 + (X_2+X_3)} + \overline{\overline{(X_0 \cdot X_3)}+ \overline{(X_1+X_2)}}\nonumber\\
    Y_2 &= (X_0 \cdot (X_1 \odot X_2)) + (X_1 \cdot (X_2 \cdot X_3))+\overline{(X_0 + X_3)} \cdot \overline{(X_1 \cdot X_2)}\nonumber\\
    Y_3 &= \overline{\overline{(X_0 \odot X_3)} + (\overline{X_1} \cdot X_2)} +\overline{(X_0 \odot X_3)} \cdot (\overline{X_1} \cdot X_2)\nonumber
    \end{aligned}\right.
\end{eqnarray}
\end{footnotesize}

The combinational implementation of {\em Spongent} Sbox represented by above equation will have $62$ fault points from $f_1, \dots, f_{62}$, where, each fault point has $3$ possible values {\em SET0}, {\em SET1}, and {\em non-faulty}. A fault combination, denoted as $\{f_i, f_j,f_k\} \rightarrow$ \{{\em SET1, SET1, SET0}\} indicates that {\em SET1, SET1, SET0} faults are injected in unison at fault points $f_i, f_j,$ and $f_k$, respectively, whereas all remaining fault-points are fault-free. Following table summarizes the observations on {\em Dumbo} cipher.

\begin{table}[!ht]
\centering
\begin{tabular}{|c|c|c|c|}
\hline
                                                             \textbf{Fault Combination}                                                            &
                                                                                                                        \textbf{Faults}
                                                                                                                                                                                                                                            &
                                                                \textbf{\#Non-occurring values at Sbox output}                                                                                                                                                                             
&

\textbf{Residual Key Space}
\\ \hline

\{$f_{13}$\} & \{SET0\} & 1 & 1   
\\ \hline 
\{$f_{13}$\} & \{SET1\} & 1 & 1   
\\ \hline

\{$f_{11}$\}& \{SET0\}                                                                                                                                                                                                          & 1                                                                                       & 1                                                                              \\ \hline
\{$f_{9}$\}                                                                                                                                                                                                                & \{SET0\}                                                                                                                                                                                                          & 1                                                                                       & 1                                                                              \\ \hline
\{$f_{14}$\},\{$f_{15}$\},\{$f_{45}$\}                                                                                                                                                                                            & \{SET1\}                                                                                                                                                                                                          & 1                                                                                       & 1                                                                              \\ \hline
\{$f_9$, $f_{38}$\}&\{SET0,SET0\} & 1                                                                                       & 1  
\\ \hline
\{$f_9$, $f_{45}$\}&\{SET0,SET0\} & 1                                                                                       & 1                                                                            \\ \hline
\{$f_{31}$\}                                                                                                                                                                                                                & \{SET0\}                                                                                                                                                                                                          & 1                                                                                       & 1 
\\ \hline

\{$f_{33}$\}                                                         & \{SET0\}                                                                                                                                                                                                          & 1                                                                                       & 1                                                                                  \\ \hline
\{$f_{33}$\}                                                &\{SET1\}& 3 & $3^{40}$\\ \hline

\{$f_14$, $f_{21}$\}&\{SET0,SET0\} & 2 & $2^{40}$                                                          \\ \hline

\{$f_15$, $f_{31}$\}&\{SET0,SET0\} & 2 & $2^{40}$                                                          \\ \hline

\{$f_14$, $f_{25}$\}&\{SET1,SET1\} & 2 & $2^{40}$                                                          \\ \hline

\{$f_15$, $f_{25}$\}&\{SET1,SET1\} & 2 & $2^{40}$                                                          \\ \hline

\{$f_31$, $f_{38}$\}&\{SET0,SET0\} & 2 & $2^{40}$                                                          \\ \hline

\{$f_31$, $f_{45}$\}&\{SET0,SET0\} & 2 & $2^{40}$                                                          \\ \hline

\{$f_33$, $f_{41}$\}&\{SET0,SET1\} & 2 & $2^{40}$                                                          \\ \hline

\{$f_25$, $f_{45}$\}&\{SET1,SET0\} & 2 & $2^{40}$                                                          \\ \hline
\end{tabular}
\caption{A few observations about {\em Single Event Transient faults} which, when subjected to the Sbox implementation lead to easy key-recovery in Dumbo cipher.}
\label{tab:faultCombinations}
\end{table}

\begin{figure}[!ht]
	\centering
	\includegraphics[width =9.5cm]{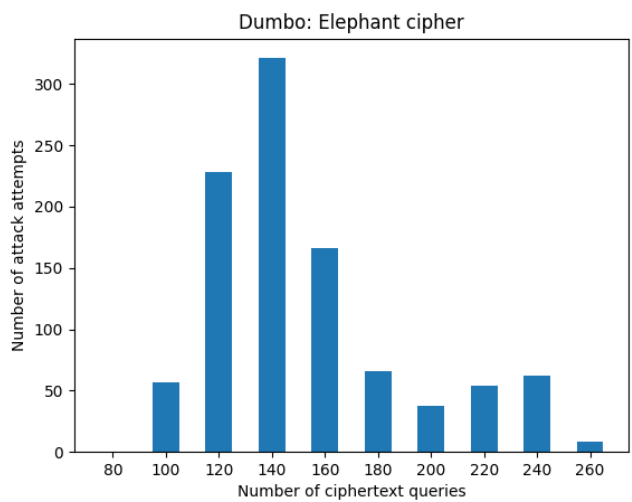}
	\caption{Success rate of proposed {\em SETFA} method. X-axis represents range of ciphertext queries (100 indicates ciphertext queries between 80-100) and Y-axis represents number of successful attack attempts(out of 1000).}
	\label{fig:successRate_elephant_isafa}
\end{figure}

We performed $1000$ independent experiments on random pairs of plaintexts, associated data, nonce, and secret key with different fault combinations of {\em Single Event Transient (SET)} faults. As a result, we found that the $160$-bit expanded key, $K'$ and corresponding $128$-bit secret key, $K$, can be recovered in $80-250$ queries for {\em Dumbo} elephant cipher. The plot shown in Fig.\ref{fig:successRate_elephant_isafa} depicts the success rate of the proposed technique on {\em Dumbo}.
 
\section{Conclusion}
In this paper, we proposed a novel fault attack called {\em Single Event Transient Fault Analysis (SETFA)}), which targets hardware implementations of cryptographic protocols with Sbox implemented as a combinational logic circuit. In addition, we proposed a full key recovery attack on NIST-LWC candidate Elephant-AEAD, which could recover the full master key in $80-250$ ciphertext queries.
\bibliographystyle{unsrt}
\bibliography{main} 

\begin{thebibliography}{1}

\bibitem{DBLP:journals/tosc/BeyneCDM20}
Tim Beyne, Yu~Long Chen, Christoph Dobraunig, and Bart Mennink.
\newblock Dumbo, jumbo, and delirium: Parallel authenticated encryption for the
  lightweight circus.
\newblock {\em {IACR} Trans. Symmetric Cryptol.}, 2020({S1}):5--30, 2020.

\bibitem{DBLP:conf/ches/BogdanovKLTVV11}
Andrey Bogdanov, Miroslav Knezevic, Gregor Leander, Deniz Toz, Kerem Varici,
  and Ingrid Verbauwhede.
\newblock spongent: {A} lightweight hash function.
\newblock In Bart Preneel and Tsuyoshi Takagi, editors, {\em Cryptographic
  Hardware and Embedded Systems - {CHES} 2011 - 13th International Workshop,
  Nara, Japan, September 28 - October 1, 2011. Proceedings}, volume 6917 of
  {\em Lecture Notes in Computer Science}, pages 312--325. Springer, 2011.

\bibitem{DBLP:conf/dagstuhl/BertoniDPA09a}
Guido Bertoni, Joan Daemen, Micha{\"{e}}l Peeters, and Gilles~Van Assche.
\newblock The road from panama to keccak via radiogat{\'{u}}n.
\newblock In Helena Handschuh, Stefan Lucks, Bart Preneel, and Phillip Rogaway,
  editors, {\em Symmetric Cryptography, 11.01. - 16.01.2009}, volume 09031 of
  {\em Dagstuhl Seminar Proceedings}. Schloss Dagstuhl - Leibniz-Zentrum
  f{\"{u}}r Informatik, Germany, 2009.

\bibitem{DBLP:conf/eurocrypt/GrangerJMN16}
Robert Granger, Philipp Jovanovic, Bart Mennink, and Samuel Neves.
\newblock Improved masking for tweakable blockciphers with applications to
  authenticated encryption.
\newblock In Marc Fischlin and Jean{-}S{\'{e}}bastien Coron, editors, {\em
  Advances in Cryptology - {EUROCRYPT} 2016 - 35th Annual International
  Conference on the Theory and Applications of Cryptographic Techniques,
  Vienna, Austria, May 8-12, 2016, Proceedings, Part {I}}, volume 9665 of {\em
  Lecture Notes in Computer Science}, pages 263--293. Springer, 2016.

\bibitem{DBLP:journals/tc/Leveugle07}
R{\'{e}}gis Leveugle.
\newblock Early analysis of fault-based attack effects in secure circuits.
\newblock {\em {IEEE} Trans. Computers}, 56(10):1431--1434, 2007.

\end{thebibliography}

\end{document}